\documentclass[runningheads]{llncs}
\usepackage{graphicx}
\usepackage{amsmath}
\begin{document}
\title{Transfer Learning by Cascaded Network to identify and classify lung nodules for cancer detection}
%
%
\author{Shah B. Shrey  \inst{1,2}
\and Lukman Hakim\inst{2}
\and Muthusubash Kavitha 
\inst{2}
\and Hae Won Kim\inst{3}
\and Takio Kurita\inst{2}
}

\authorrunning{Shrey et al.}
%
\institute{Birla Institute of Technology and Science, Pilani, India \and
Department of Information Engineering, Hiroshima University, Japan \and Department of Nuclear Medicine, Keimyung University Dongsan Medical Center, Daegu, Korea
}
\maketitle              
\begin{abstract}
Lung cancer is one of the most deadly diseases in the world. Detecting such tumors at an early stage can be a tedious task. Existing deep learning architecture for lung nodule  identification used complex architecture with large number of parameters. This study developed a cascaded architecture which can accurately segment and classify the benign or malignant lung nodules on computed tomography (CT) images. The main contribution of this study is to introduce a segmentation network where the first stage trained on a public data set can help to recognize the images which included a nodule from any data set by means of transfer learning. And the segmentation of a nodule improves the second stage to classify the nodules into benign and malignant. The proposed architecture outperformed the  conventional methods with an area under curve value of 95.67\%. The experimental results showed that the classification accuracy of 97.96\% of our proposed architecture outperformed other simple and complex architectures in classifying lung nodules for lung cancer detection. 

\end{abstract}

\section{Introduction}
Lung cancer is one of the deadliest cancers in existence. The mortality rate due to lung cancer is higher than to colorectal, breast, and prostate cancers combined [1]. Anyone can get lung cancer and approximately 60\% to 65\% of all new lung cancer diagnoses are among people who have never smoked or are former smokers ~\cite{ref_article1,ref_article2,ref_article3,ref_article4,ref_article5}. Only 19\% of all people diagnosed with lung cancer will survive 5 years or more, but if it’s caught before it spreads, the chance for 5-year survival improves dramatically~\cite{ref_article1}. The difficulty in diagnosing the lung cancer arises from the fact that it never shows the symptoms in the earlier stages.  The fact that early diagnosis can significantly improve the survival rates of the patients makes it a challenging yet important task. 

Computed tomography (CT) imaging is one of the most effective and has been widely used for the detection of lung cancers. However, non invasive methods of early stage cancer detection is important ~\cite{ref_article6}. On average the radiologists subjective measurement of  lung nodules takes around 2-3.5 minutes per slice of CT scan and also there can be variations in their judgements ~\cite{ref_article7}. Therefore, an unbiased automatic model which can quickly diagnosing the nodules for lung cancer is an important task. A lot of computer aided techniques have been attempted in the past \cite{ref_article8},\cite{Hong2008AutomaticLN}, \cite{INPROCEEDINGS7342723} but it is shown clearly that deep learning techniques~\cite{ref_article9}, \cite{MIL}, \cite{Nature}, \cite{2018}
have a significant advantage over the others~\cite{CNN}, \cite{ref_article10}. However, the lack of annotated data makes it challenging to train a deep neural networks due to their dependence on the number of data set. 
\begin{figure}[b]
\centering
\includegraphics[width=0.6\linewidth]{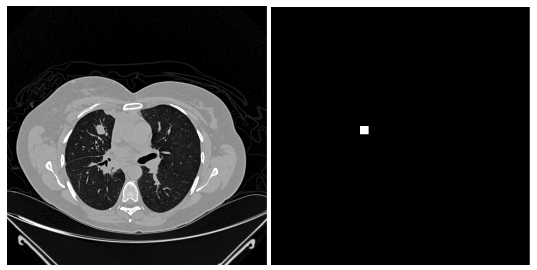}
\caption{a) Original CT image, b) generated ground truth mask.} 
\label{Mask_Generated}
\end{figure}
This study proposed to use a U-Net architecture to segment the lung nodules from the CT scan images and thus screen the CT scan slices that are suspicious of having nodules. The model employed the idea of transfer learning by training on publicly available data set and then testing on a private data set. The subsequent part of the paper compared various methods to classify the resulting CT scan images containing nodules into cancerous and non-cancerous. The performance of the proposed encoder followed by fully connected network is compared with the simple fully connected and encode-decoder followed by fully connected network in classifying the lung nodules for lung cancer detection, along with some of the other pre-existing models such as Resnet50, VGG and Densenet. 

\section{Materials and Methods}
We used the existing LUNA data set to train the model for lung nodule segmentation ~\cite{ref_url1}. The data set included 888 CT images. The LIDC/IDRI database also contains annotations which were collected during a two-phase annotation process using four experienced radiologists. Each radiologist marked lesions as they identified as non-nodule, nodule \textless 3 mm, and nodules $\ge$ 3 mm. The reference standard of our challenge consists of all nodules $\ge$ 3 mm accepted by at least 3 out of 4 radiologists. Annotations that are not included in the reference standard (non-nodules, nodules \textless 3 mm, and nodules annotated by only 1 or 2 radiologists) are referred as irrelevant findings. The ground truth mask for each of the cancerous nodules was generated by using the nodule centre and the diameter value of the image pixels indicated in the annotation file. The annotation file is stored as a csv file that contains one finding per line. Each line holds the SeriesInstanceUID of the scan, the  world coordinate pixels x, y and z position of each finding and the corresponding diameter in mm. The annotation file contains 1186 nodules. The corresponding nodule centre and diameter value for the image pixels were turned into white and the remaining pixels were turned into black as shown in Fig \ref{Mask_Generated}.

The testing was done on a private data set where the ground truth for segmentation was absent but the label for each patient whether benign or malignant was used. The data set consisted of 102 benign and 102 malignant patients with CT images.  
The private data set acquired from Keimyung University Dongsan Medical Center, South Korea. Each subject has CT volume and PET volume data set. The CT resolution was 512 x 512 pixels at 0.98mm x 0.98mm, with a slice thickness and a inter-slice distance of 3mm. The PET resolution was 128 x 128 pixels at 2.4mm x 2.4mm, with slice thickness and interslice distance of 33mm. 

\begin{figure}[t]
\centering
\includegraphics[width=\linewidth]{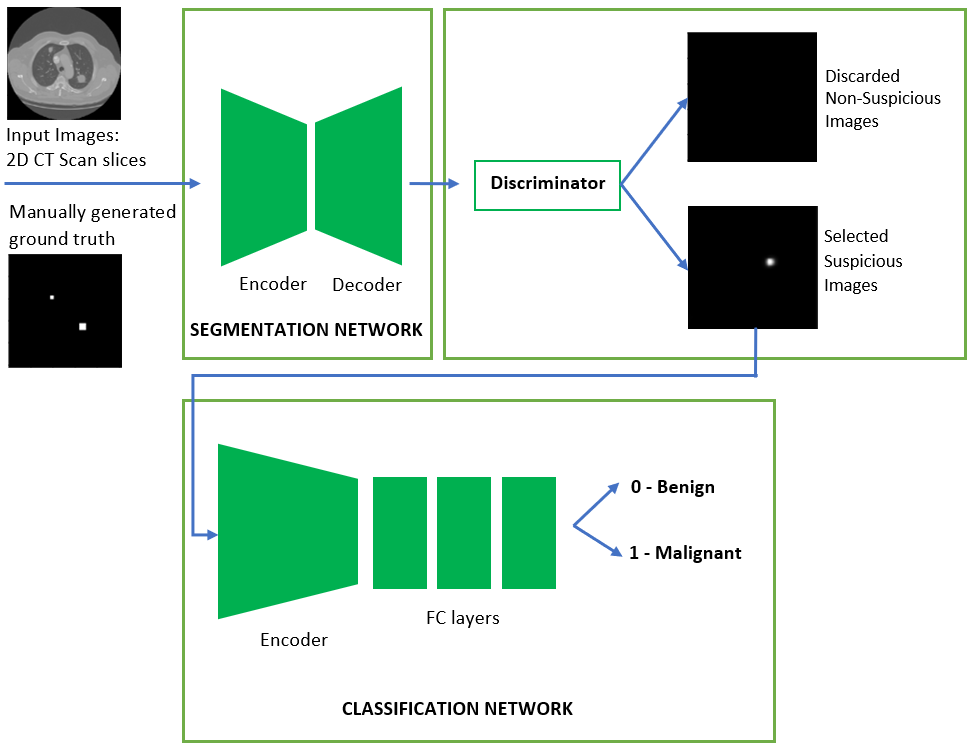}
\caption{Proposed cascaded architecture for  the identification of lung cancer.} 
\label{Overall_Architecture}
\end{figure}

\section{Cascaded Architecture}
This study proposed to suggest a cascaded network to segment and classify benign or malignant nodules for the identification of lung cancer. The segmentation network used to recognize the suspected nodules on CT images which are then classified by using classification network into benign or malignant nodules. The proposed cascaded architecture is shown in Fig \ref{Overall_Architecture}.

\subsection{Segmentation Network}

 We proposed two-stage lung cancer identification network using CT lung data sets. We used public data set of lung CT slices to train the segmentation network. The lung region slices and their corresponding  ground truth for nodules were generated as it was described earlier in the annotation file. We have included the slices between starting and ending slices along with five more slices before and after the starting and ending slice of the ground truth.

\begin{figure}
\centering
\includegraphics[width=1\linewidth]{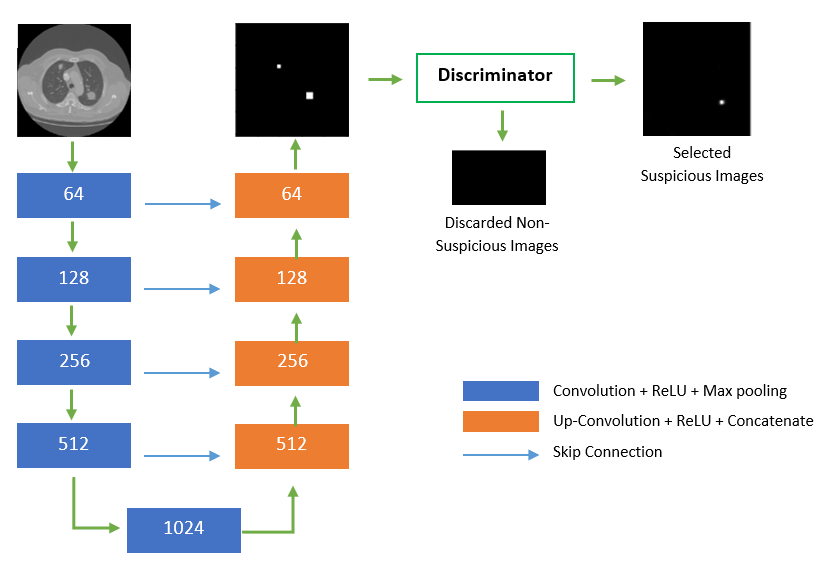}
\caption{U-Net based segmentation network to segment lung nodule for lung cancer detection} 
\label{Unet_Architecture}
\end{figure}

\begin{figure}
\begin{minipage}{0.5\textwidth}
\centering\includegraphics[width=\linewidth]{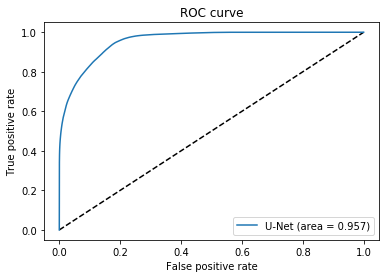}
\label{ROC_Curve}
\end{minipage}%
\begin{minipage}{0.5\textwidth}
\centering
\includegraphics[width=\linewidth]{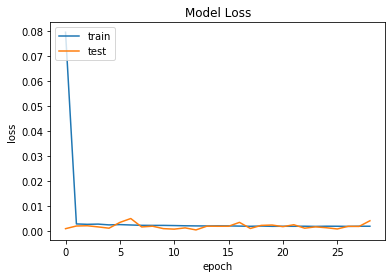}
\label{Segmentation_Loss}
\end{minipage}%
\vspace{-10pt}
\caption{Performance plot of the lung nodule segmentation. The left and right figures shows the ROC curve and loss variations, respectively for the segmentation network.} 
\label{Segmentation_ROC_Loss}
\end{figure}

A typical U-Net type model~\cite{ref_article11} was proposed as shown in Fig \ref{Unet_Architecture}. For the encoder, we used five convolutional layers, each layer with a ReLU activation. Each convolutional layer has 3 x 3 kernels and the number of channels starts from 64 and doubles every layer. A max pooling layer is applied after every convolution layer that reduces the size of the channel by half. The purpose of using pooling layers is to progressively reduce the spatial size of the representation to reduce the amount of parameters and computation in the network, and hence to also control over fitting. The Max Pooling Layer operates independently on every depth slice of the input and resizes it spatially, using the MAX operation. A 50\% drop out is applied on last two convolutions after applying ReLU activation. ReLU activation function is a piece-wise linear function which prunes the negative part to 0 and retains the positive part. It is much faster as compared to other activation functions due to it's simple max operation. For the decoder, we used four upsampling convolutions layers, each layer followed by ReLU. The output of a upsampling convolution is concatenated with an output of the corresponding part of the decoder. The softmax  with the binary cross entropy loss function is calculated for accounting the error value. The receiver operating characteristic curve (ROC) along with the loss variation based on the test data set is presented in Fig \ref{Segmentation_ROC_Loss}.

The segmentation network consists of 2 stages, the first stage is network trained  with public dataset to predict the output image pixels $y_i$ from a given input image pixels $x_i$. So, we can define the loss function on segmentation network $L_1$ as :

\begin{equation}
L_1 = \sum_i^M \{t_{i}log(y_{i})+(1-t_i)log(1-y_i)\}
\end{equation}
where $(x_i,t_i)|i=1,...,M$, $x_i$ is a $i^{th}$ input data from the training public dataset $X$, and $t_i$ is a $i^{th}$ from data target or label $T$ of public dataset. The number of training samples and labels is denoted by $M$ and $N$, respectively. 

The second stage is  trained network  used to predict the nodule segmentation from the private data set. The trained weight of the segmentation network is utilized as a screening network to find the presence of lung nodules on the private lung CT data set. 

In the segmentation architecture we used discriminator, that decides whether the segmented image consisted a nodule or not. The objective criterion used in the discriminator was if the maximum value of the pixel values $>$0.35 then the resultant image contains suspicious nodules, otherwise discarded.

\begin{figure}
\centering
\includegraphics[width=\linewidth]{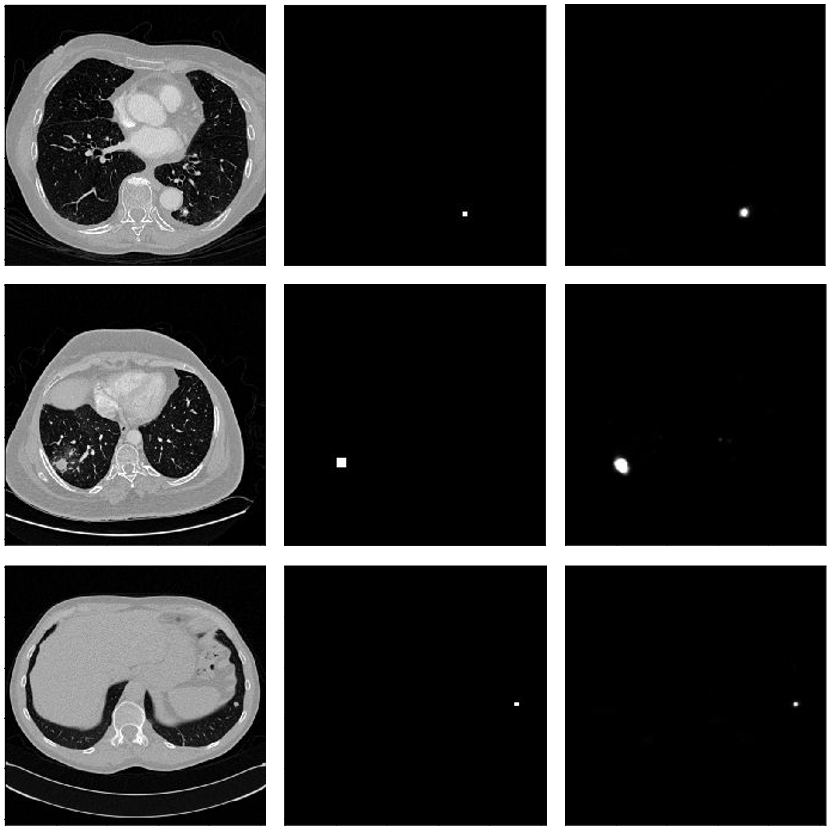}
\caption{Segmented results of lung nodules. The left image shows the original CT. The middle image is the ground truth. The right image shows the predicted lung nodule for lung cancer using our segmentation network..} 
\label{Test_on_LUNA_Data_set}
\end{figure}

\subsection{Classification Network}

\begin{figure}[!t]
\centering
\includegraphics[width=0.9\linewidth]{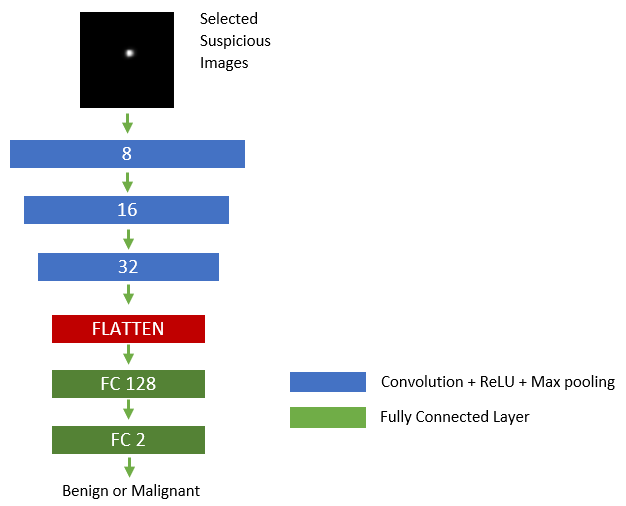}
\caption{Encoder based classification network to classify the lung nodules for lung cancer detection} 
\label{classification_network}
\end{figure}

The images which were suspected to have nodules by the previous stage segmentation network were further used in the classification network for classifying nodules. The images did not include nodules were omitted for training. 

The segmented nodule images  with their corresponding original CT images  and labels  indicating whether the nodules are benign or malignant were used as input into the coder network for classification. For this task, we define input $\bar{x_i}$ for classification network as : 

\begin{equation}
    \bar{x_i} = \{\hat{x_i},\hat{y_i}\}
\end{equation}
where $(\hat{x_i}\in \hat{X})$ is original image from private dataset and $(\hat{y_i} \in \hat{Y})$ is image with suspicious nodule. The classification network architecture is trained to predict the class labels $\Bar{y_i}$ from a given input  $\bar{x_i}$. If $(\hat{t_i} \in \hat{T})$ is labels for private dataset, we can define the loss function on classification network $L_2$ as :

\begin{equation}
L_2 = \sum_i^M \{\hat{t_{i}}log(\Bar{y_{i}})+(1-\hat{t_i})log(1-\Bar{y_{i}})\}
\end{equation}

\begin{figure}
\begin{minipage}{0.5\textwidth}
\centering
\includegraphics[width=\linewidth]{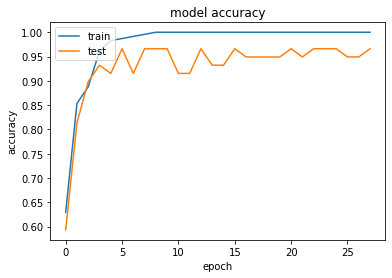}
\label{Classification_Accuracy}
\end{minipage}%
\begin{minipage}{0.5\textwidth}
\centering
\includegraphics[width=\linewidth]{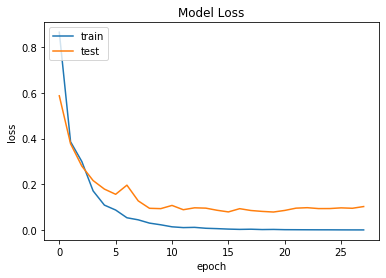}
\label{Classification_Loss}
\end{minipage}%
\vspace{-10pt}
\caption{Performance plot of the lung nodule classification. The left and right figures shows the accuracy and loss variations, respectively for the classification network.} 
\label{Classification_Loss_Accuracy}
\end{figure}

The ratio of the benign and malignant nodules is 1:5.  Therefore, to increase the number of benign nodules we used  thrice the oversampling technique to rectify the data imbalance problem. 
The encoder with the fully connected layer is developed to classify the nodules for the lung cancer. The proposed classification network is shown in Fig \ref{classification_network}. For the encoder, we used three convolutional layers, each layer with a ReLU activation. Each convolutional layer has 3 x 3 kernels and the number of channels starts from 8 and doubles every layer. A max pooling layer is applied after every convolution layer that reduces the size of the channel by half. A 50\% drop out is applied on last convolution after ReLU activation. Dropout changed the concept of learning all the weights together to learning a fraction of the weights in the network in each training iteration. Dropout is highly effective in reducing over-fitting of the network. It prevents the network from being too reliant on one or a small group of neurons, and can force the network to be more accurate even in the absence of certain information. The output is then flattened and fed into a fully connected layer with 128 nodes. It is transferred to a two node fully connected layer which is used to predict the class of the nodule in one-hot fashion. The loss and accuracy variation of our proposed classification network based on the test data set is shown in Fig \ref{Classification_Loss_Accuracy}.

\section{Experimental Settings}
\subsection{Experiment}

The U-Net model for segmentation network used  train, validation and test data set. Out of 888 CT images, the data set used for train, validation and test is 80\%, 5\% and 15\%, respectively. For segmentation network, we used  batches of 16 images and their corresponding ground truth. Adam optimizer with a learning rate of 1e-4 is used. The encoder network for classification used train, validation and test data set. For classification, the data set used for train, validation and test is 60\%, 15\% and 25\%, respectively. We used Adam as an optimizer and set the learning rate of 1e-4. All networks were implemented using Keras and Tensorflow backend with 4GB memory.

\subsection{Evaluation}
 
 We evaluated our proposed segmentation network based on U-Net with discriminator and classification network based on encoder with fully connected for the segmentation and classification of two classes of lung nodules on private CT data set. The efficiency of the architecture in predicting the two classes of lung nodules is compared to the ground truth. We compared the performance of our proposed classification network based on encoder with fully connected network with simple fully connected network and encoder and decoder followed by the fully connected layer, along with other pre-existing models like Resnet50, VGG and DenseNet. In the fully connected network the image is flatten and directly fed into the fully connected network layers. The encoder and decoder followed by the fully connected layer network is similar with the segmentation network architecture. All classification networks were trained with a similar hyper-parameter values.  The quantitative networks performance was measured using the average value of precision, recall, F1 score and accuracy. If the predicted class region belongs to the valid ground truth region, then it is considered a true positive (TP), otherwise, it is considered as a false positive (FP). If the predicted class region is correctly identified but does not belong to the valid groundtruth region; then, it is considered a true negative (TN). 
 
  We carried out additional experiments to show the ability of the proposed segmentation network with positron emission tomography (PET) images. PET images were used as a reference to confirm the reliability of our segmentation network that accurately recognizing the benign and malignant nodules.


\section{Results}

\begin{figure}
\centering
\includegraphics[width=1\linewidth]{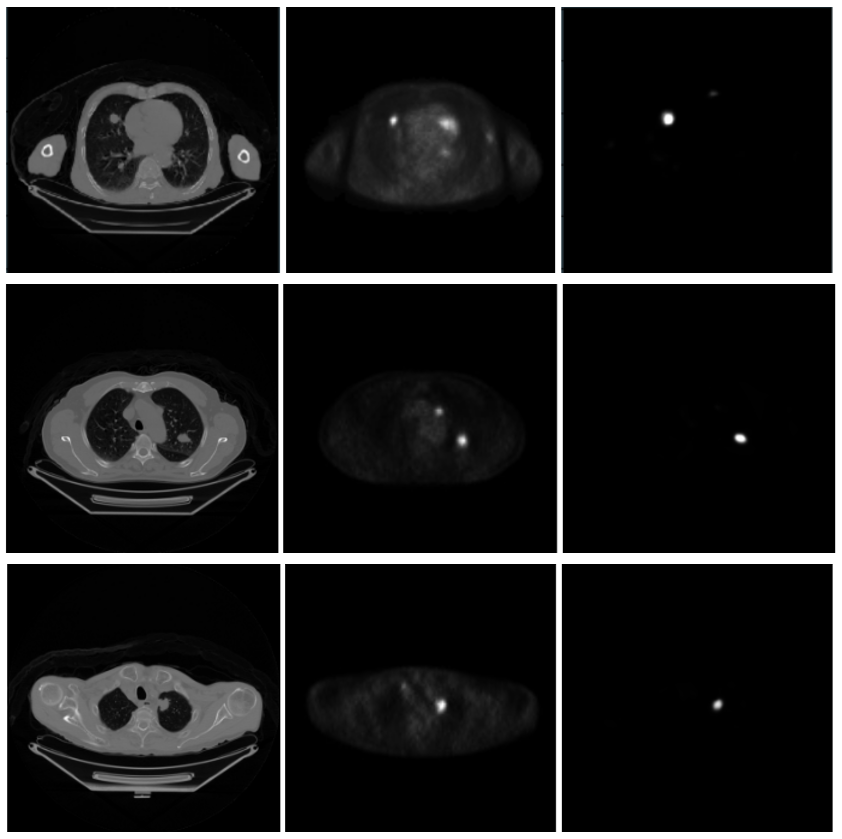} 
\caption{Segmented results of lung nodules. The left image shows the original CT.  The middle image is the reference PET image for lung nodule. The right image shows the predicted lung nodule for the original CT image on the left for lung cancer using our segmentation network.} 
\label{Test_on_Private3}
\end{figure}

The visualization of few representative examples of segmentation of nodules for lung cancer is presented in Fig \ref{Test_on_LUNA_Data_set}. In addition, the proposed segmentation network shows highest performance in recognizing the lung nodules is confirmed with the reference PET image which is shown in Fig \ref{Test_on_Private3}. 
 The comparison of  the performance of our proposed encoder with fully connected network outperforms the simple fully connected network and encoder and decoder followed by the fully connected layer. 
 
 Our proposed classification network shows highest precision, recall and F1 score of 98.0\%, and accuracy of 97.9\%, which is much higher than the coupled encoder and decoder network as shown in table \ref{Table1}. Furthermore the performance of our proposed classification network is higher than the simple fully connected network. It is because the simple fully connected network is not capable of extracting precise features for classification. In addition, our proposed model outperformed the state-of-the-arts such as Resnet50, VGG and Densenet with a huge difference of accuracy as shown in table \ref{Table2}. This can be attributed to the fact that our segmentation model was superior and boosted the performance of our classification model over the existing models. Furthermore, other existing methods for lung cancer classification performed using deep learning model produced 80.0\% \cite{Fusion} and 94.50\% \cite{Future} accuracy. Whereas, our proposed cascaded architecture for lung cancer identification achieved high accuracy of 97.9\%, indicated the effectiveness of the proposed framework in lung nodule location for cancer.

\begin{table}
\caption{Performance comparison of the proposed cascaded encoder network with different classification network  architectures for lung cancer detection}\label{Table1}
\begin{tabular*}{\textwidth}{c @{\extracolsep{\fill}} ccccc}
\hline
{\bfseries Network Architectures} & {\bfseries Precision} & {\bfseries Recall} & {\bfseries F1 Score} & {\bfseries Accuracy}\\
\hline
Fully connected & 92\% & 91\% & 91\% & 90.62\% \\
Encoder-decoder with fully connected & 26\% & 51\% & 34\% & 51.02\% \\
Proposed & {\bfseries 98\%} & {\bfseries 98\%} & {\bfseries 98\%} & {\bfseries 97.96\%} \\
\hline
\end{tabular*}
\end{table}

\begin{table}
\caption{Performance comparison of the proposed cascaded network classification with the  state-of-the-arts}\label{Table2}
\begin{tabular*}{\textwidth}{c @{\extracolsep{\fill}} ccccc}
\hline
{\bfseries Network Architectures} & {\bfseries Precision} & {\bfseries Recall} & {\bfseries F1 Score} & {\bfseries Accuracy}\\
\hline
Resnet50 & 71.25\% & 71.93\% & 71.57\% & 77.14\% \\
VGG & 75.94\% & 75.86\% & 75.33\% & 75.93\%\\
Densenet & 76.89\% & 76.47\% & 75.80\% & 76.50\%\\

Proposed & {\bfseries 98\%} & {\bfseries 98\%} & {\bfseries 98\%} & {\bfseries 97.96\%} \\
\hline
\end{tabular*}
\end{table}

\section{Conclusion}
We proposed two-stage cascaded architecture for the segmentation and classification of benign and malignant nodules for lung cancer detection. In the cascaded architecture, U-Net based segmentation network performed as a screening network and transfer the trained weights of the public data set of CT slices to the private CT slices that did not consisted ground truth for lung nodule localization. In addition the segmentation network improves the performance and robustness in classifying benign or malignant lung nodules. The experimental results suggested that our proposed encoder followed by fully connected layers classification network outperformed other classification networks for the identification of lung cancer.


\begin{thebibliography}{8}
\bibitem{ref_article1}
Howlader N, Noone AM, Krapcho M, Miller D, Bresi A, Yu M, Ruhl J, Tatalovich Z, Mariotto A, Lewis DR, Chen HS, Feuer EJ, Cronin KA (eds). SEER Cancer Statistics Review, 1975-2016, National Cancer Institute. Bethesda, MD, http://seer.cancer.gov/csr/1975\_2016, based on November 2018 SEER data submission, 2019.

\bibitem{ref_article2}
Burns, DM. Primary prevention, smoking, and smoking cessation: Implications for future trends in lung cancer prevention. Cancer, 2000; 89:2506-2509.

\bibitem{ref_article3}
Thun, MJ, et al. Lung Cancer Occurrence in Never-Smokers: An Analysis of 13 Cohorts and 22 Cancer Registry Studies. PLOS Medicine, 2008: 5(9):e185. doi: 10.1371/journal.pmed.0050185.

\bibitem{ref_article4}
Satcher D, Thompson TG, Kaplan, JP. Women and Smoking: A Report of the Surgeon General. Nicotine Tob Res, 2002; 4(1): 7-20.

\bibitem{ref_article5}
Park ER, et al. A snapshot of smokers after lung and colorectal cancer diagnosis. Cancer, 2012; 12:3153-3164. doi: 1002/cncr.26545; http://oninelibrary.wiley.com/doi/10.1002/cncr.26545/abstract.

\bibitem{ref_article6}
Diederich S., Heindel W., Beyer F., Ludwig K., Wormanns D. Detection of pulmonary nodules at multirow-detector CT: Effectiveness of double reading to improve sensitivity at standard-dose and low-dose chest CT. Eur. Radiol. 2004;15:14–22.

\bibitem{ref_article7}
Bogoni L., Ko J.P., Alpert J., Anand V., Fantauzzi J., Florin C.H., Koo C.W., Mason D., Rom W., Shiau M., et al. Impact of a computer-aided detection (CAD) system integrated into a picture archiving and communication system (PACS) on reader sensitivity and efficiency for the detection of lung nodules in thoracic CT exams. J. Digit. Imaging. 2012;25:771–781. doi: 10.1007/s10278-012-9496-0.

\bibitem{ref_article8}
Sluimer IC, van Waes PF, Viergever MA, van Ginneken B. Computeraided diagnosis in high-resolution CT of the lungs. Med Phys. 2003;30:3081–3090. doi: 10.1118/1.1624771.

\bibitem{Hong2008AutomaticLN}
Helen H, Jeongjin L, Yeny Y. Automatic lung nodule matching on sequential CT images,
Computers in biology and medicine, 2008; 38(5),623-34.

\bibitem{INPROCEEDINGS7342723} 
S. Ignatious, R. Joseph. 
Computer aided lung cancer detection system, 
Global Conference on Communication Technologies (GCCT),2015; 555-558.

 

\bibitem{ref_article9}
Greenspan H, Summers RM, van Ginneken B. Deep learning in medical imaging: overview and future promise of an exciting new technique. IEEE Trans Med Imaging. 2016;35(5):1153–1159. 

\bibitem{MIL}
Kavitha MS, Yudistira N and Kurita T, "Multi instance learning via deep CNN for multi-class recognition of Alzheimer's disease," 2019 IEEE 11th International Workshop on Computational Intelligence and Applications (IWCIA), 2019, 89-94.

\bibitem{Nature}
Ardila, D., Kiraly, A.P., Bharadwaj, S. et al.
End-to-end lung cancer screening with three-dimensional deep learning on low-dose chest computed tomography. Nat Med, 2019; 25, 954–961

\bibitem{2018} 
Tekade R, Rajeswari K,
Fourth International Conference on Computing Communication Control and Automation (ICCUBEA), 
lung cancer detection and classification using deep learning, 2018; 1-5. 

\bibitem{CNN}
Kavitha MS, Kurita T, Park SY, Chien SI, Bae JS, Ahn BC. Deep vector-based convolutional neural network approach for automatic recognition of colonies of induced pluripotent stem cells. PLoS One, 2017;12(12)

\bibitem{ref_article10}
Ginneken, B.V. Fifty years of computer analysis in chest imaging: rule-based, machine learning, deep learning. Radiol Phys Technol, 2017; 10, 23–32.

\bibitem{ref_url1}
The LUNA16 Challenge. https://luna16.grand-challenge.org/, 2016.

\bibitem{ref_article11}
Ronneberger, O., Fischer, P., \& Brox, T. U-net: Convolutional networks for biomedical image segmentation. In International Conference on Medical image computing and computer-assisted intervention, 2015; 234-241. Springer, Cham.

\bibitem{ref_article12}
Zou, KH, Warfield, S K, Bharatha, A, Tempany C. M, Kaus MR, Haker SJ, Kikinis R. Statistical validation of image segmentation quality based on a spatial overlap index. Academic radiology, 2004; 11(2), 178–189. 

\bibitem{Fusion}
Xie Y, Zhang J, Xia Y, Fulham M, Zhang Y.
Fusing texture, shape and deep model-learned information at decision level for automated classification of lung nodules on chest C.T
Inf. Fusion, 2018; 42,102-110.

\bibitem{Future}
Lakshmanaprabu SK, Sachi NandanMohanty, Shankar K.,Arunkumar N, Gustavo Ramireze (2019).
Optimal deep learning model for classification of lung cancer on CT images. Future Generation Computer Systems, 2019; 92, 374-382.






\end{thebibliography}
\end{document}